\documentclass{JAC2003}
\usepackage[usenames,dvipsnames]{color}
\definecolor{darkred}{rgb}{0.5,0,0}
\definecolor{darkgreen}{rgb}{0,0.5,0}
\definecolor{darkblue}{rgb}{0,0,0.5}
\usepackage{graphicx}
\usepackage{dcolumn}
\usepackage{bm}
\usepackage{epsfig}
\usepackage{enumerate}
\usepackage[a4paper]{hyperref}
\hypersetup{ colorlinks,
linkcolor=darkred,
filecolor=darkgreen,
urlcolor=darkblue,
citecolor=darkred }

\usepackage{epstopdf}
\newcommand{\be}{\begin{equation}}
\newcommand{\ee}{\end{equation}}

\title{ELECTRON CLOUD IN THE CERN ACCELERATORS (PS, SPS, LHC)}
\author{G.~Iadarola$^{(1+2)}$ and G.~Rumolo$^{(1)}$\\
$^{(1)}$ CERN, Geneva, Switzerland, $^{(2)}$ Universit\`{a} di Napoli ``Federico II'', Naples, Italy}

\begin{document}
\maketitle

\abstract
Several indicators have pointed to the presence of an Electron Cloud (EC) in some of the CERN accelerators, when operating with closely spaced bunched beams. 
In particular, spurious signals on the pick ups used for beam detection,
pressure rise and beam instabilities were observed at the Proton Synchrotron (PS) during the last stage of preparation of the beams for the Large Hadron Collider (LHC),
as well as at the Super Proton Synchrotron (SPS). Since the LHC has started operation in 2009, typical electron cloud phenomena have appeared also
in this machine, when running with trains of closely packed bunches (i.e.~with spacings below 150ns).
Beside the above mentioned indicators, other typical signatures were seen in this machine (due to its operation mode and/or more refined detection possibilities), 
like heat load in the cold dipoles, bunch dependent emittance growth and degraded lifetime in store and bunch-by-bunch stable phase shift to compensate for the energy
loss due to the electron cloud.\\
An overview of the electron cloud status in the different CERN machines (PS, SPS, LHC) will be presented in this paper, with a special emphasis on the dangers for
future operation with more intense beams and the necessary countermeasures to mitigate or suppress the effect.

\vspace*{0.1cm}
\section{Introduction}\vspace{-0.cm}
In the CERN PS, the electron cloud was first observed in 2001 during the last part of the cycle for the production of the the so-called LHC-type beams, 
i.e. the beams of the type needed for the LHC filling. The production scheme of these beams in the PS is based on two or three steps of bunch splitting
in order to obtain at the exit of the PS bunch trains with 50ns or 25ns spacing, respectively. In either case, the final stage of bunch splitting takes place at the 
top energy (26 GeV/c) and is followed by adiabatic bunch shortening and fast bunch rotation shortly before extraction
\cite{heiko}. These two processes are meant to shorten the bunches from their 15~ns length after the last 
splitting to 12 and then 4~ns, respectively, and make them suitable to be injected into the SPS. Therefore, these beams only circulate in the PS for few tens of msec with a structure prone to electron cloud formation (beam parameters are summarized in 
Table \ref{PS-beam}).\\ 
During this short time before extraction, an electron cloud was initially revealed in 2001 by the presence of a baseline drift in the signal from the pick up as well as 
beam transverse instabilities \cite{cappi}. The transverse instabilities made a new appearance with 25ns beams in 2006, when the bunches were accidentally shortened 
to 10ns or below (instead of the nominal 12ns) during the phase of adiabatic shortening prior to the fast rotation. This again suggested that the short bunches could
initiate the electron cloud build up earlier in the cycle and produce enough electron cloud for a sufficiently long time as to render the beam visibly unstable. In
March 2007, an experiment for dedicated electron cloud measurements was set up at the PS to be able to directly measure the electron signal by using a shielded
biased pick up \cite{mahner} and confirm its presence in the machine in the last phase of the LHC beams production. The experimental setup was designed, fabricated, 
and mounted in straight section (SS) 98 during the accelerator shutdown 2006/2007. All the details of the setup can be found in Ref.~\cite{mahner}. These studies
confirmed that the electron cloud develops during the last 40 to 50 ms before ejection, i.e.~when the bunches are shortened by the RF gymnastics. Besides, they also
showed that the electron cloud can be suppressed by putting a sufficiently large voltage of either polarity onto a clearing electrode, even if the clearing efficiency
depends on the magnetic field present in the region of the measurement in a non-trivial way.
\vspace*{-0.4cm}

\begin{table}[ht]
\begin{center}
\caption{{\small Relevant beam parameters in the PS during the flat top RF gymnastics for the two bunch spacings of 50 and 25ns.}}
\vspace{0.2cm}
\begin{tabular}{|c|c|c|}
\hline
 & {\bf 50ns} & {\bf 25ns} \\ \hline\hline
Beam energy (GeV/c) & \multicolumn{2}{c|}{26} \\ \hline
 \begin{tabular}{c}Bunch intensity\\ ($\times 10^{11}$ ppb)\end{tabular} & 0.82-1.95 & 0.83-1.33  \\ \hline
Bunch length (ns) & \multicolumn{2}{c|}{15 $\rightarrow$ 12 $\rightarrow$ 4} \\ \hline
Number of bunches & 36 & 72 \\ \hline
Transv.~norm.~emittances ($\mu$m) & 1-2 & 2-3 \\ \hline
\end{tabular}
\label{PS-beam}
\end{center}
\end{table}
\vspace*{-0.2cm}
In 2011, new systematic measurements of electron cloud have been performed at the CERN-PS with the goal of extracting the following information:  
\begin{itemize}
\item Dependence of the electron cloud build up evolution on some controllable beam parameters (bunch spacing, bunch intensity, bunch length).
\item A new collection of time resolved experimental data of electron cloud build up in some desired sets of beam conditions. 
\end{itemize}
These sets of data can serve two purposes. First, comparing them with 
build up simulations will allow us to validate (or improve) the simulation model on which our tools are based. Second, by matching the simulations to the experimental 
data in all the different beam conditions, we can pin down the surface properties of the PS vacuum chamber (secondary electron yield, $\delta_{max}$, and reflectivity of 
the electrons at zero energy, $R_0$) and extrapolate then how much electron cloud we can expect in the PS with the higher intensity beams foreseen in the frame of the 
LHC Injector Upgrade (LIU) project, and whether that can be detrimental to the beam.
\begin{table}[ht]
\begin{center}
\caption{{\small Relevant beam parameters of the SPS 50 and 25ns beams.}}
\vspace{0.2cm}
\begin{tabular}{|c|c|c|}
\hline
 & {\bf 50ns} & {\bf 25ns} \\ \hline\hline
Beam energy (GeV/c) & \multicolumn{2}{c|}{26 $\rightarrow$ 450} \\ \hline
 \begin{tabular}{c}Bunch intensity\\ ($\times 10^{11}$ ppb)\end{tabular} & 0.3-1.7 & 0.3-1.4  \\ \hline
Bunch length (ns) & \multicolumn{2}{c|}{4 $\rightarrow$ 2.8 $\rightarrow$ 1.6} \\ \hline
Number of bunches & 144 & 288 \\ \hline
Transv.~norm.~emittances ($\mu$m) & 1-2 & 2-3 \\ \hline
\end{tabular}
\label{SPS-beam}
\end{center}
\end{table}
\vspace*{-0.2cm}

The SPS has been suffering from electron cloud formation since it first began to take and accelerate 25ns beams produced in the PS with the scheme explained above. 
Observations of pressure rise, beam instability, emittance growth were first made in the early 2000 and all these effects strongly limited the capability of this 
accelerator of handling LHC-type beams \cite{arduiniPAC2011}. While the coherent instabilities could be suppressed by the use of the transverse damper (against the
horizontal coupled bunch oscillations) and running with sufficiently high chromaticity (against the strong single bunch effect in the vertical plane), emittance
growth and positive tune shift along the bunch train could still be measured, pointing to the continuing presence of a strong electron cloud inside the beam chamber. 
All this led to the decision to have in 2002 the first dedicated scrubbing run, in which the SPS was operated exclusively
with 25ns beams for one full week. The goal was to use the bombardment from the electron cloud itself to clean the beam chamber inner surface, and therefore lower its 
Secondary Electron Yield (SEY) and reduce, in turn, the amount of electron cloud build up. The strategy proved successful \cite{jimenez2003} and the week of scrubbing
run was then repeated at the beginning of the 2003, 2004, 2006 and 2007 runs to provide the necessary machine cleaning. During these years, dedicated experiments were conducted in 
the SPS to study in detail
the electron cloud formation in cold regions (COLDEX) or in NEG coated chambers \cite{baglinrossi-ECLOUD04}, or to benchmark simulation codes with machine observations 
\cite{schultePAC2005}. From 2006 on, electron cloud studies in the SPS acquired new momentum in the framework of the SPS upgrade studies \cite{valencia} and the
experimental activity over the following years was mainly focused to find the scaling law of the electron cloud instability with beam energy \cite{rumoloPRL} and to
validate the efficiency of amorphous carbon (a-C) coating of the beam chamber \cite{yinvalgren}.\\
All the electron cloud machine development activity of the last couple of years at the SPS has been devoted to defining the status of the 25ns beams in this machine 
and use the direct electron cloud measurements in chambers equipped with strip monitors to understand beam induced scrubbing in different chamber geometries and with
different materials. A comprehensive report of all observations in terms of beam behavior, pressure rise and dedicated ekectron cloud measurements will be soon published
\cite{barto-iada}. Recently, the nominal 25ns and 50ns LHC beams in the SPS seem to be undegraded and does not suffer any longer from the strong electron cloud
effect that was present during the first years of SPS operation with this type of beams. The achievable parameters are summarized in Table \ref{SPS-beam}. The three values of bunch length quoted in this table correspond to injection into 
2~MV buckets, after shortening at flat bottom by increase of the RF voltage to 3~MV, and at flat top after controlled longitudinal emittance blow up during the
accelerating ramp.\\
Several studies conducted in the past predicted that also 
the LHC would suffer from heat load, pressure rise and beam instabilities due
to electron cloud, when operating with trains made of closely spaced proton bunches (e.g. \cite{zimmermann-chamonix}). 
Since mid 2010 LHC entered this mode of operation.
In the first phase, beams with 150~ns bunch spacing were injected, accelerated and brought to collision. During this period of
operation, the only possible signature of electron cloud build up was a pressure rise observed in the common vacuum
chamber, close to the Interaction Regions. Subsequently, at the end of October 2010, an attempt was made to switch to 50~ns 
spacing operation. After an initial physics fill with 108 nominal bunches (filling scheme with 1 pilot 
bunch and $9\times 12$ bunches), some important dynamic pressure rises were observed at injection when filling with trains 
of 24 bunches. In fact, the first attempt of injection in batches of 24 even led to the closure of 
the vacuum valves in point 7 after the injection of 108 nominal bunches per beam, as the interlock level of 
$4\times 10^{-7}$~mbar was reached on two vacuum gauges. After that, since it became clear that further improvements
in the LHC performance were hampered by the electron cloud, emphasis was put on machine studies
to characterize the electron cloud build-up in the LHC, its effects and possible cures. It was also decided that a comparative 
study with the behaviour of 75~ns beams was necessary to define a path for the 2011 run. Toward the end of the 2010 proton 
run, a Machine Development (MD) session was devoted to the set up of the LHC with 50~ns bunch trains.
During this MD, three effective days of beam time
were used for the setting-up proper as well as for studies 
and machine scrubbing. The study of the 75~ns beam took place in another dedicated MD period, while the 
LHC had already switched to ion operation. About 2.5 days were devoted to 
the setting-up of the injection and capture of the 
75~ns beam and, later on, to comparative studies with the 50~ns beam. This MD gave a clear indication that, probably also
benefiting from the previous MD's scrubbing with 50~ns beams, the electron cloud effects with 75~ns appeared 
significantly less pronounced than with 50~ns beams, such that this bunch spacing could be regarded as a 
relatively safe option \cite{gianluigi}.

\begin{figure*}[tb]
\begin{center}
\rotatebox{0}{\scalebox{0.67}{\includegraphics{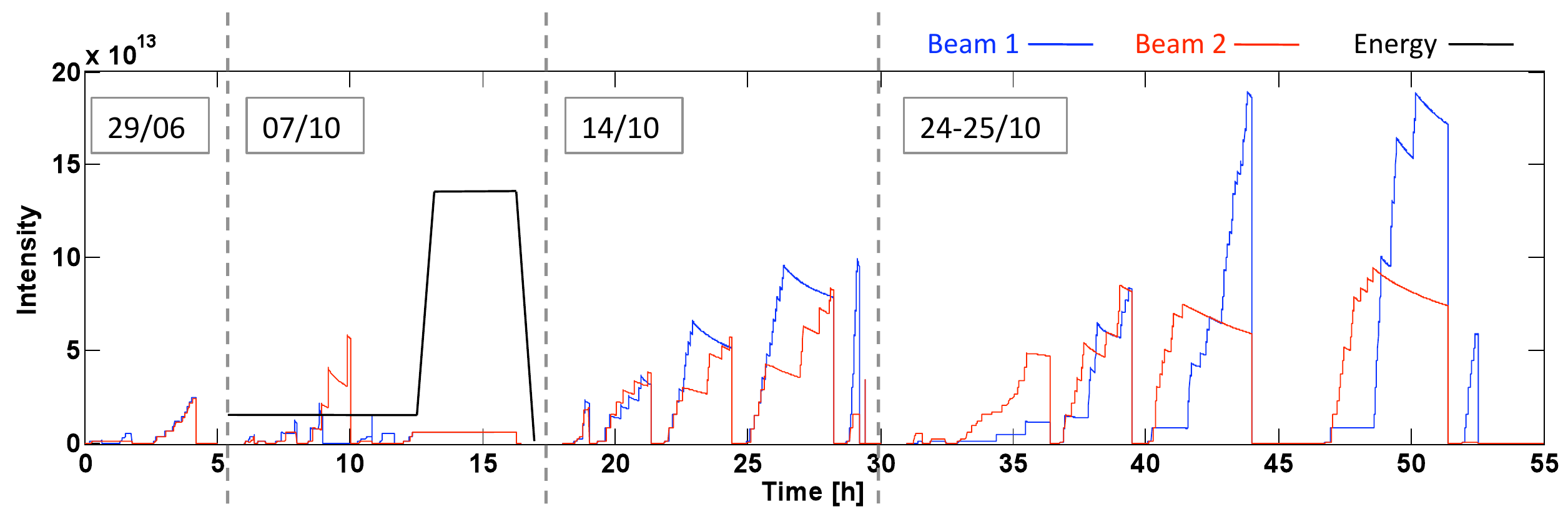}}} 
\vspace*{-0.4cm}
\caption{{\small MD sessions labeled (a), (c), (d) and (e): injected beams.
}}
\label{GRfig1}
\end{center}
\vspace*{-0.4cm}
\end{figure*}

The LHC operation was therefore resumed in 2011 directly with 75~ns beams. After the scrubbing run in 2010 it was expected that up 
to 200-300 bunches could be injected and accelerated without major problems. This was confirmed during the start-up with beam.
After about one month of operation, the LHC could successfully collide trains of 200 bunches 
distributed in batches of 24 bunches each. At the beginning 
of April, 10 days were devoted to scrubbing of the LHC with 50~ns beams. The goal was to prepare the machine to switch to 50~ns 
beams and thus extend the luminosity reach for the 2011 run. During the scrubbing run, up to 1020 bunches per beam were injected
into the LHC in batches of 36 and stored at injection energy. The strategy consisted of constantly topping the total beam 
intensity in the LHC
with the injection of more trains, such that the vacuum activity, and therefore the electron cloud, could be kept at a constant 
level and efficiently reduce the Secondary Electron Yield (SEY) of the walls to a value below the threshold for build up. The
success of the scrubbing run was proved by the subsequent smooth LHC physics operation with 50~ns spaced beams. Between mid 
April and end June the number of bunches collided in the LHC was increased up to 
its maximum value of 1380 per beam, while the intensity per bunch and the transverse emittances remained constant at their
nominal values (i.e., 1.15$\times 10^{11}$ ppb and 2.5~$\mu$m). The switch to 50~ns beams with 
lower transverse emittances (1.5~$\mu$m) and the adiabatic increae of the bunch current to  $1.5\times 10^{11}$~ppb did not cause any
significant recrudescence of the electron cloud effects, probably also owing to the MD sessions with 25ns beams that took place
in the second half of 2011, which created enough margin in the machine cnnditioning to ensure electron cloud free 50ns operation.\\
Beams with 25ns spacing were injected into the LHC only during five MD sessions of the 2011 run, which are listed and briefly described 
here:
\begin{enumerate}[(a)]
\item {\bf 29 June, 2011}: first injections of 25ns beams into the LHC. The filling scheme consisted of nine batches of 24 bunches separated by 
increasing gaps (2.28, 5.13 and 29.93 $\mu$s). Pressure rise around the machine as well as heat loads in the arcs were observed. All the
last bunches of each batch suffered losses and emittance growth \cite{brennan};
\item {\bf 26 August, 2011}: first injections of a 48-bunch train into the LHC with 25ns spacing. Two attempts were made to inject a 48-bunch 
train from the SPS, which led to beam dump triggered by large beam excursion and beam loss interlocks, respectively. During the first 
injection test, the transverse damper was on and it is believed that the beam suffered a coherent electron cloud instability in both planes 
(more critical in vertical) soon after injection. During the second test, the transverse damper was switched off and the beam was affected by
a coupled bunch instability \cite{hanneswolfgang}. This MD session had then to be 
interrupted because of a cryo failure caused by a thunderstorm;
\item {\bf 7 October, 2011}: injection tests and first ramp. In the first part of the MD, trains with 48-72-144-216-288 bunches from 
the SPS were injected into the LHC. Given the experience during the previous MD, the chromaticity $Q'$ was set to around 15-20 units in both 
the horizontal and vertical planes in order to keep the beams stable against the electron cloud effect. In the second part, only 60 bunches 
per beam were injected in trains of $12 + 2\times 24$, were accelerated to 3.5~TeV and collided during approximately 5h;
\item {\bf 14 October, 2011}: first long stores of 25ns beams at injection energy in the LHC. During this session up to 1020 bunches per beam
were injected in batches of 72. The chromaticity was kept high in both planes ($Q'_{x,y}\approx 15$) in order to preserve the beam
stability. First, a dedicated fill for pressure measurements was made, with batches injected at gradually reduced distances 
from 4 to 2~$\mu$s (in steps of 1$\mu$s).
Subsequently, the batch spacing was kept constant for each of the next three fills and it was set to 6.3, 3.6 and 1~$\mu$s (rounded values).
Strong emittance growth and slow losses affecting the last bunches of each train were observed throughout this MD session;
\item {\bf 24--25 October, 2011}: record number of bunches in the LHC. Four long fills took place (average store time was approximately 4h), 
with 25ns beams injected into both rings in batches of 72 separated by 1$\mu$s. In the third and fourth fills, 2100 bunches were injected
for beam 1, while the number of bunches could not exceed 1020 for beam 2, due to a vacuum interlock on one of the injection kickers
(MKI). Although the situation seemed to improve over the MD, slow losses and emittance growth kept affecting both beams. Before starting the 
fourth fill, the horizontal chromaticity $Q'_x$ was lowered from 15 to 3 units and the horizontal damper gain was slightly increased. Probably
due to that, some horizontal instabilities could be observed from the signal of the damper pick up during the fourth fill, but the overall 
performance did not appear degraded from the previous fill. The MD ended with a 30' fill with only beam 1, during which batches of 72 bunches 
were injected into the LHC at different spacings in order to provide the stable pressure measurements needed for the modeling of the electron 
cloud build up in the straight sections (see next Section).
\end{enumerate}

Figure \ref{GRfig1} shows the detailed story, in terms of injected beams 1 \& 2, of the sessions (a), (c), (d) and (e). Experimental data from these
MDs will be used in the next section to extrapolate the evolution of $\delta_{\mathrm{max}}$ on the beam screen in the arcs and in proximity
of the vacuum gauges. For sake of compactness, we have chosen to concatenate these three sessions and represent them as a function of a 
continuous time coordinate (interpretable as hours with 25ns beam), which will be systematically used throughout this paper 
when referring to the studies with the 25ns beams. 

\section{Studies in the different machines}
\subsection{PS measurements}
\begin{figure}[htb]
\begin{center}
\rotatebox{0}{\scalebox{0.55}{\includegraphics{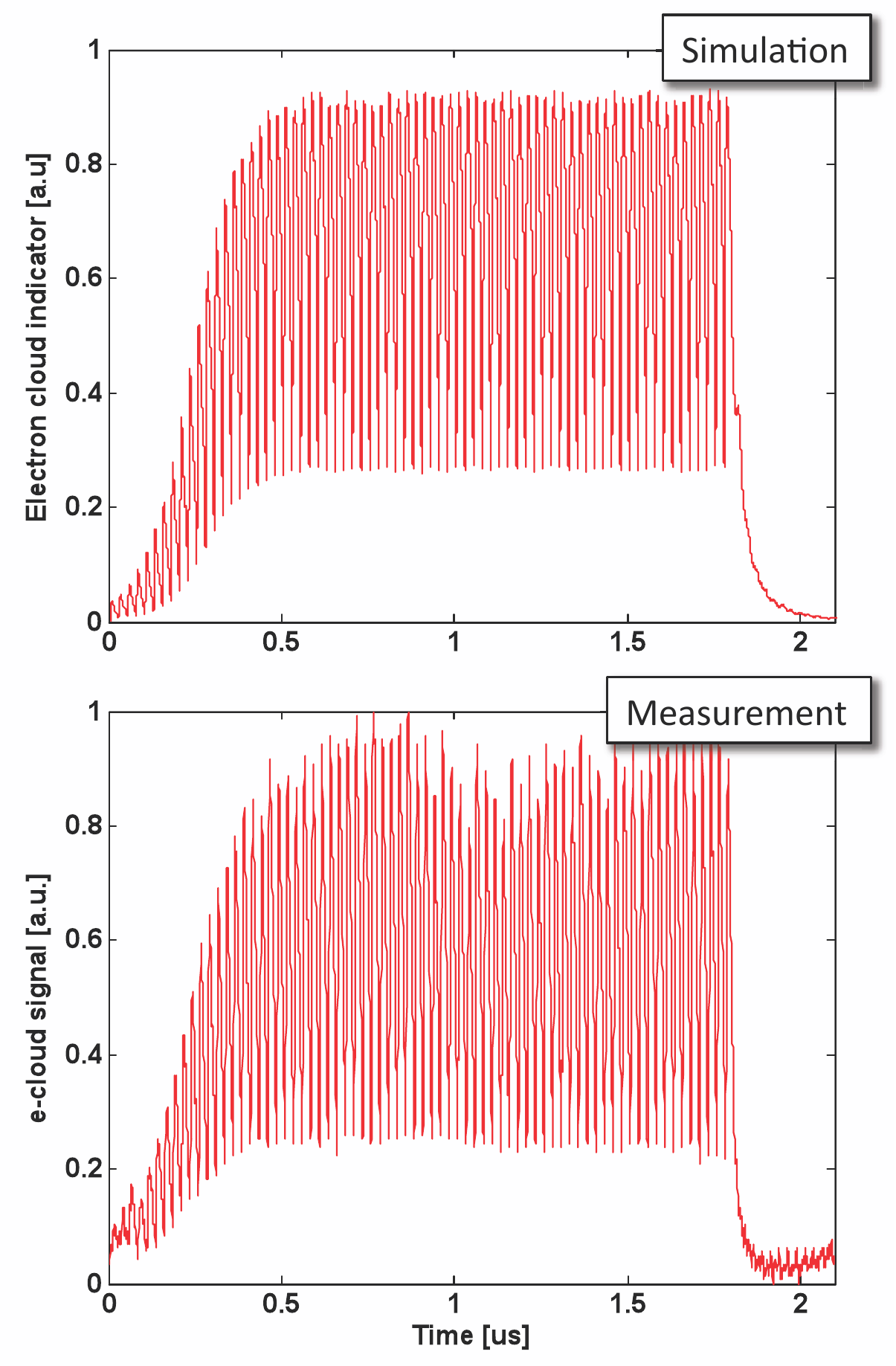}}} 
\vspace*{-0.2cm}
\caption{{\small E-cloud build up simulation (top) and measurement (bottom) for a 25ns beam with $1.33\times 10^{11}$~ppb and 4ns long.
}}
\label{fig5}
\end{center}
\vspace*{-0.5cm}
\end{figure}

In 2011, the MD program in the PS for electron cloud studies took place in November and extended over several sessions to cover different sets 
of beam
parameters. In particular, electron cloud build up data were recorded for 25ns and 50ns beams. The bunch intensities were scanned in the ranges indicated in Table
\ref{PS-beam}. The trigger for the data acquisition was set at extraction, when in normal conditions each bunch of the beam has been already fully
rotated (4ns bunch length). However, specifically for these measurements, the bunch length at this time for a fixed bunch intensity was also set to 
6.5ns or 15ns by simply adjusting or fully removing, respectively, the final step of the fast bunch rotation. This allowed studying the dependence
of the electron cloud build up not only on the bunch intensity but also on the bunch length.\\
The threshold for electron cloud formation with 50ns beams was found to lie
at about $10^{11}$~ppb and the measured signal increases monotonically with the bunch intensity. This is not entirely surprising, since the measurements were taken with 
zero magnetic field while the non-monotonic behaviour of the electron cloud build up with the bunch intensity is more frequent in dipole regions. 
The shielded pick up is installed inside a C-magnet, which was kept off during the MD sessions because the orbit perturbation it introduces
would have required a specific correction. Scans with 25ns beams were also made and the threshold for electron cloud formation
was found to be below $8\times 10^{10}$~ppb, with a behavior of the electron cloud signal increasing with the bunch intensity.\\
We have tried to fit the PS data with those from electron cloud build up simulations \cite{iadarola}. First of all, the output of the code that
should be compared with the measured signal is the electron flux to the wall. In a first approximation, we do not consider the holes in the vacuum chamber, which
are expected to cause only a minor perturbation in a field-free region. In general, the simulated electron flux to the wall vanishes during the bunch
passage, because initially all the electrons are drawn to the center of the vacuum chamber by the passing bunch (e.g.~during the first {\small $\sim$}2ns of a 4ns
long bunch) and they are gradually
released only during the falling edge of the bunch, when they may reach the walls again. The fact that the measured signal does not exhibit this feature
makes plausible a low pass filtering of the signal (inherent to the measurement technique or due to electronics and/or cables) with a corner
frequency in the range of some hundreds of MHz. Figure \ref{fig5} shows measured and simulated signal, where
the simulated signal, obtained with $\delta_{\mathrm{max}}=1.6$ and $R_0=0.5$, was low pass filtered with a corner frequency of 200~MHz. The impressive resemblance between
the two suggests that our electron cloud model correctly describes the phenomenon and the rationale applied for the data analysis is promising.

It is therefore clear that the electron cloud is present in the CERN PS with both 50ns and 25ns beams when they reach the final beam
structure, shorty before being ejected. However, since it only makes a short appearance in the last few ms of the production cycle of these beams,
with the present beam parameters, there is not enough time to render beam unstable or let incoherent effects develop. On the negative side,
very low electron doses are deposited on the chamber walls, making it basically impossible to rely on efficient machine scrubbing if the electron
cloud should ever become a limiting factor. The question to be addressed is whether this effect may become a bottleneck for the LHC Injector Upgrade (LIU) 
beams, envisaging bunch intensities of up to $3\times 10^{11}$~ppb within lower transverse emittances. A full simulation study including both the build up
and instability part is needed to assess the margins.
\begin{figure}[tb]
\begin{center}
\rotatebox{0}{\scalebox{0.66}{\includegraphics{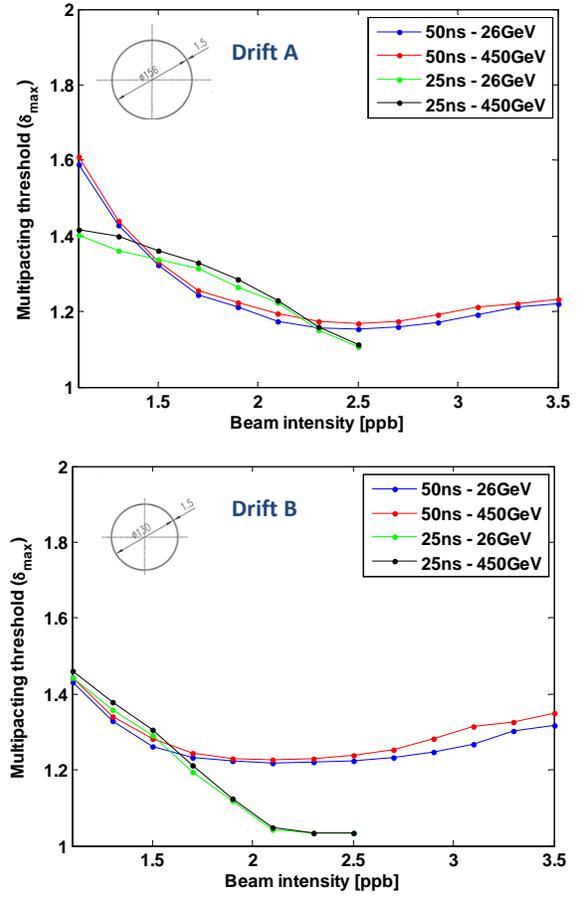}}} 
\vspace*{-0.2cm}
\caption{{\small Threshold SEY for electron cloud formation in the two types of chambers in SPS drift spaces, as a function of the bunch intensity.
}}
\label{GRfigSPSdr}
\end{center}
\vspace*{-0.5cm}
\end{figure}
\subsection{SPS studies}
\begin{figure}[tb]
\begin{center}
\rotatebox{0}{\scalebox{0.66}{\includegraphics{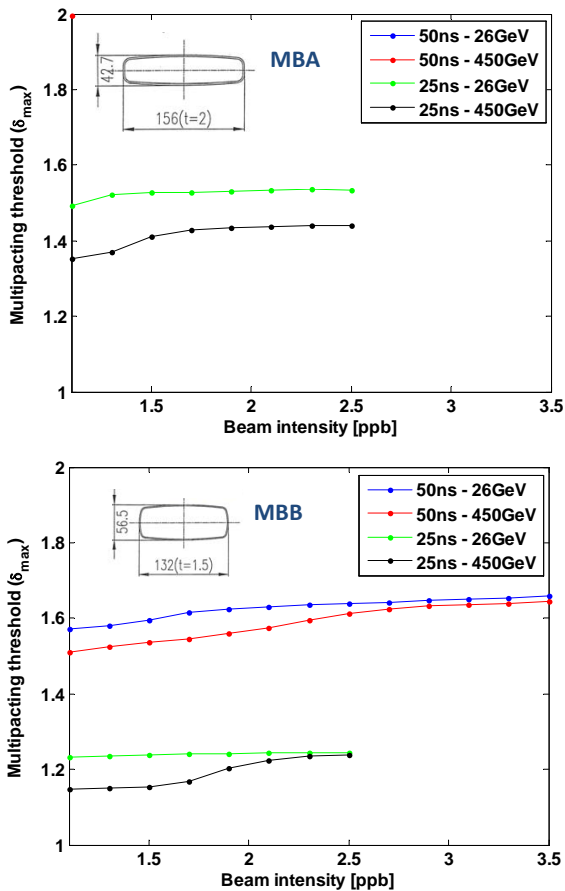}}} 
\vspace*{-0.2cm}
\caption{{\small Threshold SEY for electron cloud formation in the two types of chambers in SPS dipoles, as a function of the bunch intensity.
}}
\label{GRfigSPSdip}
\end{center}
\vspace*{-0.5cm}
\end{figure}

One of the key points to be addressed to understand the electron cloud in the SPS is to determine the values of SEY thresholds for its 
formation in the different beam chambers and try to deduce what parts are critical for both present and future LHC beams. In the SPS there are six main different
types of vacuum chamber: two types are used in the main dipoles, two in the quadrupoles and two in the drift spaces, depending on the beta functions in the
nominal optics. We have studied the electron cloud build up in both dipole chambers and driift space chambers. Since the magnetic fields from quadrupoles have not been
implemented yet in the PyECLOUD code, these chambers, which however cover only less than 10\% of the total circumference, have not yet been simulated.
The drift chambers are of A or B type, both circular and with a radius of 78 or 65~mm, respectively. Figure \ref{GRfigSPSdr} shows the SEY threshold as function of
bunch intensity at both 26 and 450~GeV/c. The following interesting features can be observed:
\begin{itemize}
\item The SEY thresholds are mostly decreasing with bunch current, but tend to change slope for 50ns beams with bunch populations above $2 \times 10^{11}$ ppb. 
\item There exist regions in which 50ns can create a worse electron cloud than 25ns.
\item The SEY thresholds become very low (close to 1.05) for 25ns beams in Drift B and with bunch currents above $2 \times 10^{11}$ ppb.
\end{itemize}
The vacuum chambers in dipoles also come in two different sorts with almost rectangular shape: the MBB-type, characterized by a height of about 56.5mm and 132mm 
width; and the MBA-type, flatter than the MBBs and thus more suited to regions with lower vertical betatron functions, characterized by a height of 43mm and 156mm width.
In Fig.~\ref{GRfigSPSdip} the SEY threshold is displayed as a function of the bunch intensity for both 25 and 50ns beams as well as at injection and 
top energy. Also in this case the dependencies are not trivial and exhibit the following features:
\begin{itemize}
\item The SEY thresholds do mostly increase with bunch current. When they do not, the behavior tends to be flat, indicsting then little dependence of the SEY threshold
on the bunch intensity in these intensity ranges.
\item The SEY thresholds of the 50ns beam lie above 2.0 in the MBA chambers.
\item The SEY thresholds can become in general very low (around 1.2) for 25ns beams in MBB chambers.
\end{itemize}

Considering all the results of the above study, it is evident that the most dangerous chambers in the SPS, in terms of favoring electron cloud build up for the
present and future LHC beam intensities, are the drift B and the MBB pipes, which exhibit the lowest SEY thresholds at almost all intensity ranges. In particular,
it is specially worrisome that both these chambers exhibit SEY thresholds below 1.3, which seems the saturation value for scrubbing of StSt in laboratory
measurements \cite{christhesis}. Besides, StSt samples exposed to the SPS beam and then extracted from the machine have never shown SEY values below 1.5. Presently,
it still remains unclear whether we still have electron cloud in some of the SPS regions, because the observed pressure rise is several order of magnitude lower than
the one observed in previous years and the nominal 25ns beam is not really affected anymore by significant electron cloud effects \cite{barto-iada}. While more
studies are ongoing to try to characterize the present status of the SPS and draw conclusions on future strategies against electron cloud, it is however clear that
the critical regions that might need coating (if scrubbing is insufficient or too long) would amount to about 40\% of the whole machine (Drift B + MBB).\\
Experimentally, we can say that, thanks to the regular scrubbing runs the SPS had from 2003 to 2008 with 25ns beams at every start up (plus several MD sessions with 
this type of beams every year), the performance with 25ns beams has been constantly improving over the years and in 2011, nominal 25ns beams with transverse emittances 
below 3$\mu$m were first produced and extracted. This leads us to believe that presently the electron cloud has weakened or disappeared in most parts of the SPS and 
might be still only surviving in the MBBs for operation with nominal intensity 25ns beams. In these conditions, it seems to be efficiently kept under control and does
not give rise to detrimental effects on the beam. An Increase in bunch intensity may awaken the electron cloud in the Drifts (and MBAs, because the stripes move to 
unscrubbed regions) with the consequent effect on beam stability and emittance evolution. It is clear that a scrubbing run will be necessary after the Long Shutdown 
2013-2014, but its length and efficiency are difficult to estimate. The experience after LS1 will therefore give an indication whether we really need to coat the most
critical parts of the SPS, or we can afford to rely on scrubbing also for future operation.

\begin{figure*}[tb]
\begin{center}
\rotatebox{0}{\scalebox{0.66}{\includegraphics{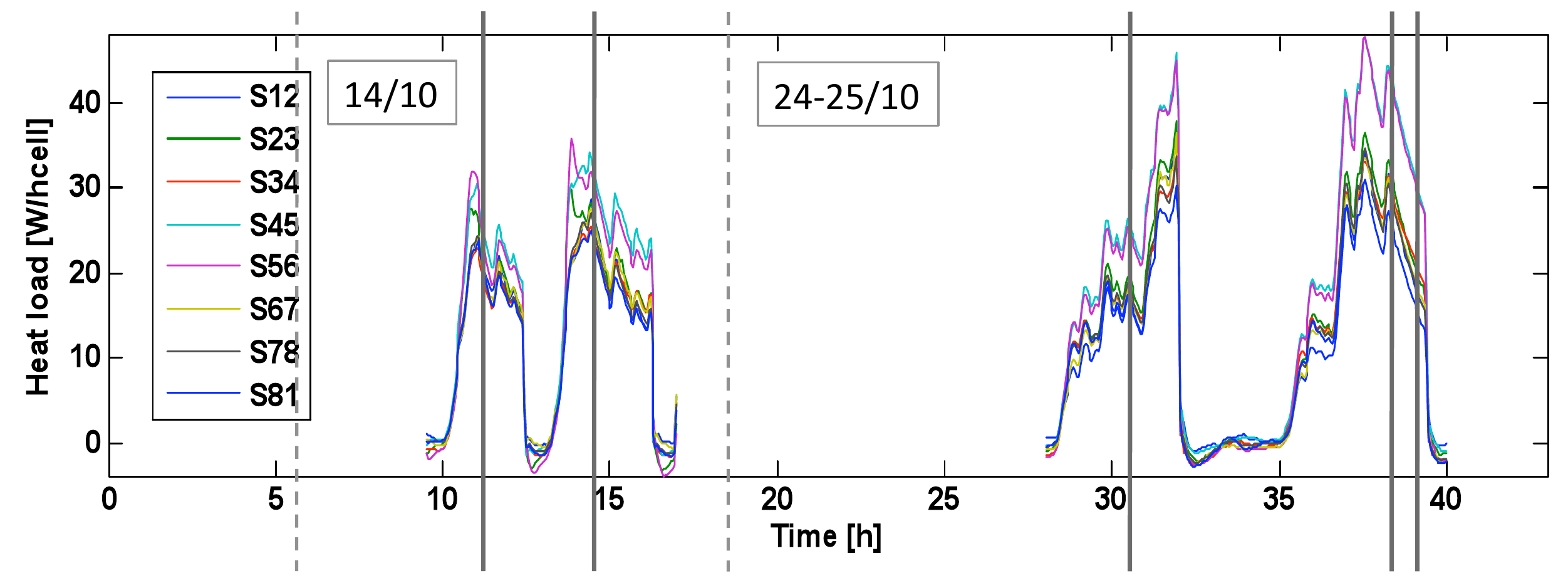}}} 
\rotatebox{0}{\scalebox{0.67}{\includegraphics{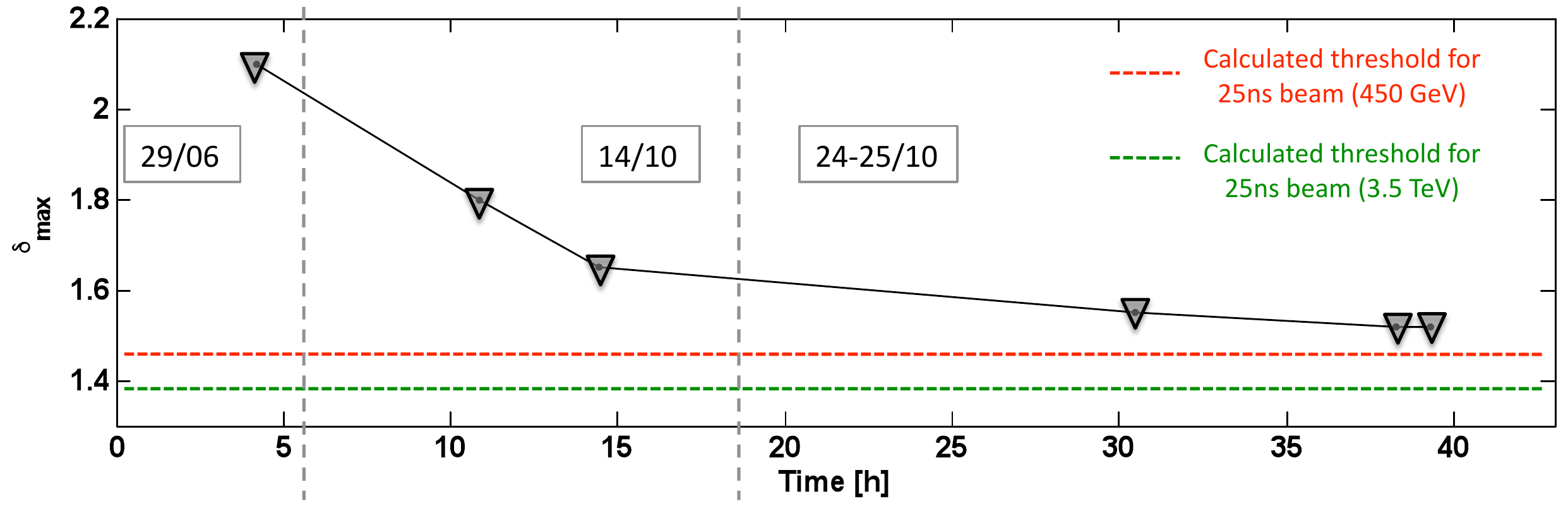}}} 
\vspace*{-0.2cm}
\caption{{\small Top picture: Heat load measured during four fills from the MD session (d) and (e), in the same time coordinate as in Fig.\ref{GRfig1}. The
five vertical bars represent the measurement points used to compare heat load with electron cloud simulations. Bottom picture: 
Estimated evolution of $\delta_{\mathrm{max}}$ on the inner surface of the beam screen in the dipole chambers
}}
\label{GRfig3}
\end{center}
\vspace*{-0.5cm}
\end{figure*}

\subsection{LHC observations} 
The heat load data from the cryogenic system give the total power dissipated (in W/half-cell) on the beam screens of both beams 1 and 2. Using the measured 
heat load it is possible to estimate the SEY of the arc chamber walls. The exact procedure is explained in Ref.~\cite{GR-evian} and is based on the comparison 
of the heat load data with PyECLOUD simulations, run with realistic bunch-by-bunch intensities and lengths (data from the fast BCT and the
BQM). Heat load observations in the arcs were made with 50ns before the scrubbing tun and then with 25ns beams. From the heat load data with 50ns beams before
and after the scrubbing run, we could estimate the SEY on the chamber wall of the arcs to have reached a value between 2.1 and 2.2, sufficient to suppress
electron cloud build up with 50ns beams. 
Measurements in some reference cells from the first LHC MD with 25ns beams (MD session (a), 29 June, 2011) can be found 
in Ref.~\cite{brennan}. Figure \ref{GRfig3} shows the heat load data, sector by sector, collected during
the MD sessions (d) and (e). We can notice that the additional heat load peaked to values of nearly
50~W/half-cell (i.e. approximately an average of 0.5~W/m/beam) during the last fill with 2100 bunches for beam 1 and 1020 bunches for beam 2.
A decay of the measured heat load between injections, and in any case after the last injection, is also clearly visible in the examined cases,
due to the weakening of the electron cloud activity from scrubbing and also from intensity loss (e.g., compare with the BCT
signal in Fig.~\ref{GRfig1}, acquired at the same time).\\
Using the bunch-by-bunch intensity and length data at the times marked
with vertical bars in the top plot of Fig.~\ref{GRfig3} plus the data from the injection tests on the 29 June, PyECLOUD simulations were run scanning 
$\delta_{\mathrm{max}}$, so that the curves of the 
simulated heat loads as a function of  $\delta_{\mathrm{max}}$ could be produced for all these
measurement points. The electron reflectivity at zero energy was fixed to the value of 0.7. The
$\delta_{\mathrm{max}}$ corresponding to each heat load measurement was then found matching the simulation to the measured value and the results
are in the curve displayed in the bottom part of Fig.~\ref{GRfig3}.\\
\begin{figure}[htb]
\begin{center}
\rotatebox{0}{\scalebox{0.67}{\includegraphics{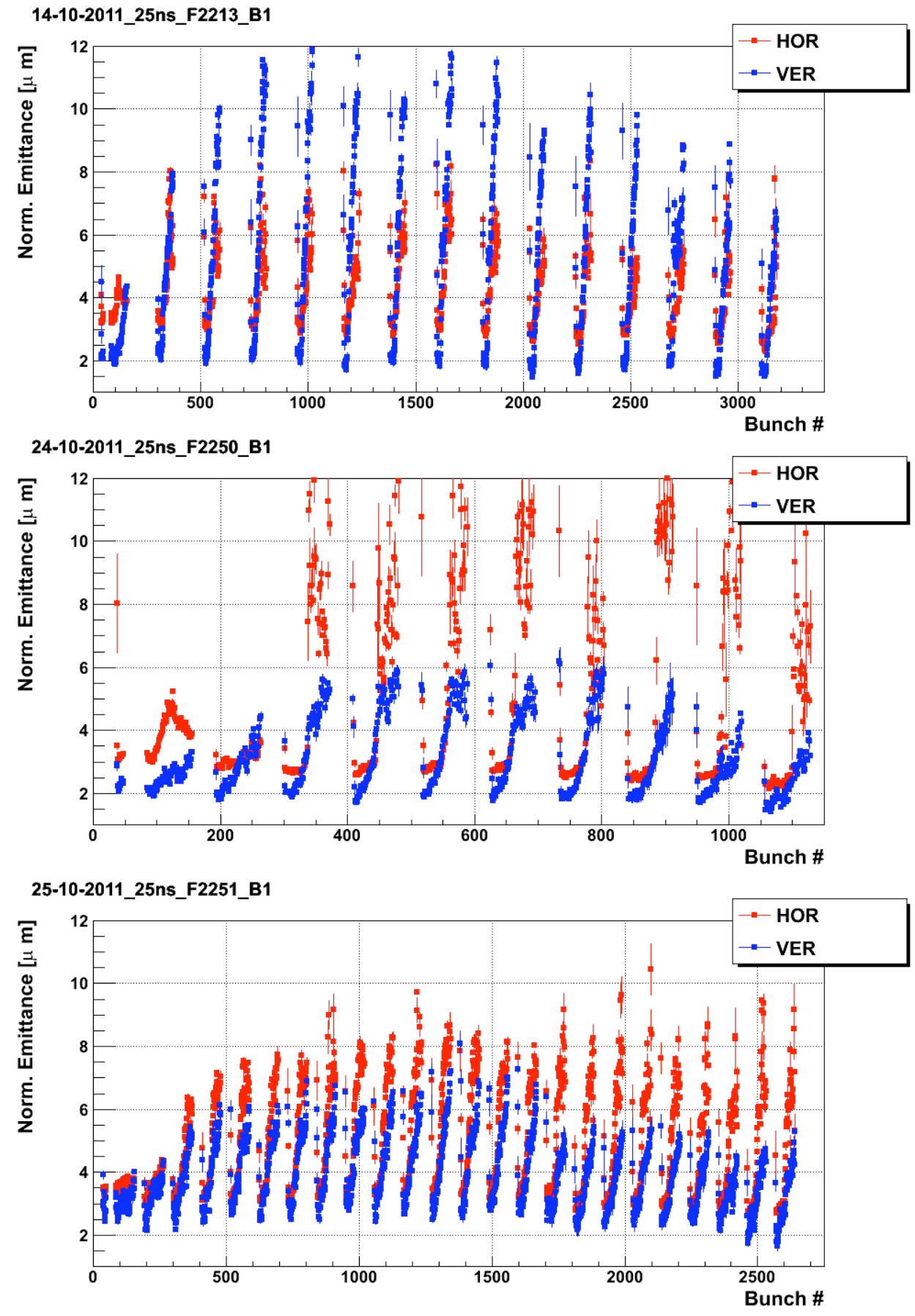}}} 
\caption{{\small Snapshots of the horizontal and vertical emittance measurements for beam 1 during the last fill of 14 October and 
the last two fills of 24--25 October MDs.
}}
\label{GRfig8}
\end{center}
\vspace*{-0.5cm}
\end{figure}
While the 50ns beam proved to be stabilized in the LHC by the electron cloud mitigation achieved with the scrubbing run, the 25ns beam has 
exhibited clear signs of transverse instability and emittance growth throughout all the dedicated MD sessions. Despite a clearly improving 
trend from one fill to the next one, these signs have not completely disappeared. During the first tests on 29 June, when 
only batches of 24 bunches were injected from the SPS, the beam could be kept inside the machine because the level of electron cloud 
reached along each batch was enough to cause significant emittance growth, but no coherent instability and fast beam loss \cite{brennan}. When,
on the following MD session, batches of 48 bunches were for the first time transferred from the SPS to the LHC, the beam was twice dumped after
few hundreds of turns, due to the excitation of a transverse instability leading to unacceptable beam losses. During the successive MD sessions,
this problem was circumvented by injecting the beam into the LHC with high chromaticity settings. Values of $Q'_{x,y}$ around 15 were chosen, as
they had been found to be sufficiently stabilizing in HEADTAIL simulations \cite{kevinli}. Using these settings, the beam could be kept inside
the LHC, albeit with degraded transverse emittances (see bunch-by-bunch emittance plots from the MDs of 
14, 24 and 25 October, Fig.~\ref{GRfig8}). Since the BSRT needs
about 2~sec to measure the emittances of each bunch, each of the snapshots in the figure does not represent an instantaneous photograph of the beam 
at a certain time, but results from a sweep over the bunches that can last as much as several minutes. Although the batch spacing was decreased
from 2~$\mu$s during the measurement of 14 October to the 1~$\mu$s of the last MD session, the vertical emittance blow up 
exhibits signs of improvement. No significant further change is observed then in the vertical plane between the measurements taken 
in the last two fills (consistently with a slight scrubbing effect between them). The situation looks more complicated in the horizontal
plane. Here a deterioration can be noticed from the 14/10 measurement to the 24/10 one. If this is related solely to the decreased batch spacing,
which has enhanced the electron cloud along the full train owing to the stronger memory effect between batches, we could not explain why we
observed an improvement in the vertical plane, instead. It is interesting that the situation
appears improved for the 25/10 measurement, when the LHC was run with lowered horizontal chromaticity settings. This fact may suggest that by
lowering chromaticity we have moved from a regime of strong incoherent emittance growth driven by electron cloud and high chromaticity to a 
new one, in which the beam suffers 
a fast instability, but later evolves with a better lifetime 
\cite{benedetto,ohmi-zim}. In any case, as a general consideration, a clear
weakening of the electron cloud effect from 14 to 25 October is witnessed by the improved quality of the first two--three
batches. The first two seem to be hardly affected by emittance growth in both transverse planes by the time of the last 25ns fill.

\section{Conclusions and outlook}\vspace{-0.cm}
In conclusion, we have reached quite a deep knowledge of the electron cloud in the different CERN accelerators, and presently it does not seem to
be a limiting factor with the present operation parameters:
\begin{itemize}
\item In the {\bf PS} the electron cloud only appears in the last milliseconds of the LHC beam production cycle and  does not stay long enough as to affect the beam
\item The {\bf SPS} currently benefits from several years of scrubbing with nominal 25ns beams. Therefore, it seems that now the electron cloud
has been either suppressed in the whole machine or it still survives in some more sensitive parts, but at a level not harmful to the beams (50ns, nominal 25ns)  
\item In the {\bf LHC}, the electron cloud does not have important adverse effects on operational 50ns beams, however it still affects the 25ns beams and
additional scrubbing is needed to further lower it and permit operation with this type of beams.
\end{itemize}
However, some questions are still open, like the performance of SPS and LHC with 25ns beams after LS 2013-2014 and whether the electron cloud can
become a serious bottleneck for the beams required by the LIU project.


\begin{thebibliography}{99}
\bibitem{heiko} R.~Garoby, ``Status of the Nominal Proton Beam for LHC in the PS'', CERN/PS 99-13 (RF)
\bibitem{cappi} R.~Cappi, {\em et al.}, Phys.~Rev.~ST Accel.~Beams {\bf 5}, 094401 (2002).
\bibitem{mahner} E.~Mahner, T.~Kroyer and F.~Caspers, Phys.~Rev.~ST Accel.~Beams {\bf 11}, 094401 (2008)
\bibitem{arduiniPAC2011} G.~Arduini, K.~Cornelis, W.~Hoefle, G.~Rumolo, and F.~Zimmermann, in Proceedings of PAC 2001 (18-23 June 2001, Chicago, USA) and CERN-SL-2001-0050
\bibitem{jimenez2003} J.M.~Jimenez {\em et al.}, LHC-Project-Report-632 (2003)
\bibitem{baglinrossi-ECLOUD04} V.~Baglin, A.~Rossi, {\em et al.}~in Proceedings of ECLOUD04 (19-23 April 2004, Napa California, USA)
\bibitem{schultePAC2005} D.~Schulte, G.~Arduini, V.~Baglin, J.M.~Jimenez, F.~Ruggiero, and F.~Zimmermann , in Proceedings of PAC2005 (16-20 May 2005, Knoxville 
Tennessee, USA) and LHC Project Report 847
\bibitem{valencia} G.~Rumolo, E.~M\'{e}tral and E.~Shaposhnikova, in Proceedings of LHC LUMI 2006 (16-20 October, Valencia, Spain)
\bibitem{rumoloPRL} G.~Rumolo, G.~Arduini, E.~M\'{e}tral, E.~Shaposhnikova, E.~Benedetto, R.~Calaga, G.~Papotti and B.~Salvant, 
Phys.~Rev.~Letters {\bf 100} (2008) 144801 
\bibitem{yinvalgren} C.~Yin Vallgren {\em et al.}, Phys.~Rev.~ST Accel.~Beams {\bf 14}, 071001 
\bibitem{barto-iada} H.~Bartosik, G.~Iadarola {\em et al.}, to be published
\bibitem{zimmermann-chamonix} F.~Zimmermann, in Proceedings of Chamonix X \& XI, CERN-SL-2000-001 DI (2000) and. CERN-SL-2001-003 DI (2001)
\bibitem{gianluigi} G. Arduini {\em et al.}, CERN-ATS-Note-2011-046 MD (2011)
\bibitem{brennan} B.~Goddard {\em et al.}, \href{http://cdsweb.cern.ch/record/1364233?ln=en}{CERN-ATS-Note-2011-050 MD} (2011)
\bibitem{GR-evian} G.~Rumolo {\em et al.}, in Proceedings of the \href{https://indico.cern.ch/internalPage.py?pageId=1&confId=155520}{LHC 
Beam Operation Workshop - Evian 2011} (12-14 December, 2011, Evian, France) 
\bibitem{hanneswolfgang} H.~Bartosik and W.~H\"{o}fle,  \href{http://cdsweb.cern.ch/record/1426592?ln=en}{CERN-ATS-Note-2012-027 MD} (2012)
\bibitem{iadarola} G.~Iadarola and G.~Rumolo, ``Improved electron cloud build up modeling with PyECLOUD'', elsewhere in these proceedings
\bibitem{christhesis} C. Yin Vallgren, Ph.D.~thesis, \href{http://cdsweb.cern.ch/record/1374938?ln=en}{CERN-THESIS-2011-063} (2011)
\bibitem{kevinli} K. Li and G.~Rumolo, \href{http://accelconf.web.cern.ch/AccelConf/IPAC2011/papers/mops069.pdf}{MOPS069} 
in proceedings of IPAC'11 (San Sebastian, Spain)
\bibitem{benedetto} E.~Benedetto, G.~Franchetti, and F.~Zimmermann, Phys.~Rev.~Lett.~{\bf 97} (2006) 034801
\bibitem{ohmi-zim} K.~Ohmi and F.~Zimmermann, Phys.~Rev.~Lett.~{\bf 85} (2000) 3821-3824
\end{thebibliography}
\end{document}